\documentclass{mn2e}

\usepackage[dvips]{graphicx}
\def\etal{{\rm et al. }}
\begin{document}

\title[AGNs and galaxy interactions]
{AGNs and galaxy interactions }

\author[M. Sol Alonso et al.]{M. Sol Alonso $^{1,2}$,
Diego G. Lambas $^{1,4}$,
Patricia Tissera $^{1,3}$ and
Georgina Coldwell $^{1,4}$\\
$^1$ Consejo Nacional de Investigaciones Cient\'{\i}ficas
y T\'ecnicas.\\
$^{2}$ Complejo Astron\'omico El Leoncito, Argentina \\
$^{3}$ Instituto de Astronom\'{\i}a
y F\'{\i}sica del Espacio, Argentina.\\
$^4$ Observatorio Astron\'omico
de la Universidad Nacional de C\'ordoba,  Argentina.\\
}

\maketitle

\begin{abstract}

We perform a statistical analysis of AGN host
characteristics and nuclear activity for AGNs in pairs and without
companions.
Our study concerns a sample of AGNs derived from the SDSS-DR4 data by
Kauffmann et al (2003) and pair galaxies obtained from the same data set
by Alonso et al. (2006).
An eye-ball classification of images of 1607 close pairs
($r_p<25$ kpc $h^{-1}$, $\Delta V<350$ km $s^{-1}$) according to the
evidence of interaction through distorted morphologies and tidal
features provides us with a more confident assessment of galaxy
interactions from this sample.
We notice that, at a given luminosity or stellar mass content, the
fraction of AGNs is larger for pair galaxies exhibiting evidence for
strong interaction and tidal features which also show sings of strong
star formation activity.
Nevertheless, this process accounts only for a $\sim 10\%$ increase of the
fraction of AGNs.
As in previous works, we find AGN hosts to be redder and with a larger
concentration morphological index than non-AGN galaxies. This effect does
not depend
whether AGN hosts are in pairs or in isolation. The OIII luminosity of
AGNs with strong
interaction features is found to be significantly larger than that of
other AGNs,
either in pairs or in isolation.
Estimations of the accretion rate, $L[OIII]/M_{BH}$, show that AGNs in
merging pairs are actively feeding their black holes, regardless of their
stellar masses.
We also find that the luminosity of the companion galaxy seems to be a key
parameter in the determination of the black hole activity.
At a given host luminosity, both the OIII luminosity and the $L[OIII]/M_{BH}$
are significantly larger in AGNs with a bright companion ($M_r < -20$)
than otherwise.

\end{abstract}


\section{Introduction}

Galaxy-galaxy interactions can be an effective mechanism to
regulate the star formation activity in galaxies.
Studies of galaxies in pairs in larger surveys such as the 2dFGRS and the
SDSS showed with high statistical signal that star formation is enhanced
by a factor
of two in close systems (Lambas et al. 2003; Alonso et al 2004).
The trend of increasing star formation for
decreasing relative velocity and projected separation was first showed by
Barton \& Geller (1998).
This trend was confirmed to be in place independently of the
characteristics of environment by Alonso et al. (2006).

The physics behind the triggering of star formation activity during
galaxy-galaxy interactions have been explained by theoretical (Martinet
1995 and
references there in) and numerical analysis
(e.g. Toomre \& Toomre 1978; Barnes \& Hernquist 1992, 1996;
Mihos \& Hernquist 1996). These studies showed that the starbursts are
fueled by gas inflows produced by the tidal torques generating during the
encounters.
The efficiency of this mechanisms depends on the particular internal
characteristics of galaxies and their gas reservoir.

Besides feeding star formation, these gas inflows could also feed a
central black hole and AGN activity (Sanders et al. 1988). The
co-evolution of galaxies
and black holes is now widely accepted although many details on how
this coexistence works are still understudied (Heckmann et al. 2004).
Toomre \& Toomre (1972) suggested that collisional disruption and gas
dissipation could feed the nuclear activity of galaxies.
Other ideas about the fueling of active galaxy nucleus are the disk of barred
galaxies generated by internal instabilities which can lead the gas to the
centre of galaxies (Schwartz 1981, Shlosman, Begelman \& Frank 1990)
and the presence of more that one supermassive black hole which seems
likely in the nucleus of merger remnants (Begelman, Blanford and Rees 1980).

Certainly, all these processes may provide clues to solve the AGN fueling
problem.
However, AGN-galaxy interaction paradigm has been the issue of several
studies which aim at determining, for example, the frequency of AGNs in a
sample of
interacting galaxies, the proximity effects of a close companion on the
AGN power, etc.
Dahari (1985), in a sample of 167 systems, found that there is an excess
of Seyferts among spiral interacting galaxies and also that Seyfert
activity and the
existence of external tidal forces are interrelated. However, Seyfert
nuclei were
not found in interacting ellipticals.
From statistical study of companions of Seyfert galaxies Fuentes-Williams \&
Stocke (1988) derived that Seyfert galaxies have a marginal excess of
comparable size companions and a clear
excess of faint companions. However, Keel (1996) found that the presence
of AGN does not seem to be related with the type of collisional interaction.

Conversely Schmitt (2001) analysed a sample of different types of active
galaxies such as LINERS, transition galaxies, absorption-line galaxies,
Seyfert and
HII galaxies. This author showed that galaxies of different activity types
have
similar percentage of companions than galaxies of similar morphological types
with no nuclear activity. This result suggests that
interactions between galaxies are not a necessary condition to trigger the
nuclear activity in AGNs.
Similar results were found by Kelm, Focardi \& Zitelli (2004) by studying
a sample of Seyfert in UZC-Compact Groups and by Coldwell \& Lambas (2006)
who
analyzed the percentage of close companions of SDSS quasars within the range
$r_p < 100 $ kpc $h^{-1}$ and $\Delta V < 350 {\rm km ~ s^{-1}}$.

An interesting approach to study the role of interactions on the feeding
of both star formation and black holes is to study AGN activity in galaxy
pairs.
In fact, Sortchi-Bergmann et al. (2001) found a correlation between
active galactic nucleus and star formation activity in interacting
galaxies by
analyzing 35 Seyfert 2 nuclei.
The galaxy spectral information available in large surveys provides unique
information and high statistical numbers
to study on more robust basis the relation between these processes.
In this paper, we combine the galaxy pair sample constructed by Alonso et
al. (2006) from the Sloan Digital Sky Survey (SDSS-DR4) with a
sample of $~ 88000$ narrow-line AGN selected by Kauffmann et al (2003,
hereafter K03) from the SDSS-DR4, which is hitherto the largest sample of
AGNs.

In section 2, we briefly describe the AGN and pair catalogs.
Section 3 analyses the properties of hosts of Type-2 AGN galaxies.
Section 4 provides a comparative statistical analysis of AGN hosts in
pairs and without companions with the aim of unveiling the role of
interactions in triggering AGN activity. In section 5 we give our main
conclusions and discussion of the results.

\section{SDSS-DR4 data}

The SDSS (York et al. 2000) in five optical bands will
map one-quarter of the entire sky and perform a redshift survey of
galaxies, quasars and stars.
The fourth data release, DR4, is the major catalogue which provides images,
spectra, and redshifts for download. The imaging portion comprises
6670 square degrees of sky imaged in five wave-bands ($u$, $g$, $r$, $i$
and $z$) containing photometric parameters of 180 million unique objects.
The main galaxy sample is essentially a magnitude limited spectroscopic
sample (Petrosian magnitude) \textit{$r_{lim}$}$ < 17.77$, most of galaxies
span a redshift range $0 < z < 0.25$ with a mean redshift at $z=0.1$
(Strauss et al. 2002).

The SDSS-DR4 has more than 500000 galaxy spectra and includes different
galaxy properties such as magnitudes, star formation indicator, concentration
index parameters, etc.
From SDSS-DR4, K03 have constructed a catalogue with a
subset of 88178 narrow emission line galaxies that are classified as AGN,
within redshift range 0.02 $ < z < $ 0.3.

Considering the unified model (Antonucci 1993), AGNs can be separated into
two categories: Type 1 AGN where the black hole and the associated continuum
and broad emission-line region is viewed directly and Type 2 AGNs where
only the narrow line region can be observed due to
the obscuring medium.
The Baldwin, Phillips \& Terlevich (BPT, 1981) line-ratio diagram allow us to
distinguish type 2 AGNs from normal star forming galaxies by considering the
intensity ratios of two pairs of relatively strong emission lines.
The sample of Type 2 AGN of K03 was selected taking into
account the relation between spectral lines, $[OIII]\lambda 5007$, $H\beta$,
$[NII]\lambda 6583$ and $H\alpha$ luminosities where an AGN is defined if

\begin{equation}
log([OIII]/H\beta) > 0.61/(log([NII/H\alpha])-0.05)+1.3
\end{equation}

We then use $~ 88000$ Type 2 AGNs in K03 catalog. The luminosity of
$[OIII]\lambda 5007$ emission line will be used as
a tracer of AGN activity.

\subsection{SSDS-DR4 galaxy pair catalog}

In our previous works (Lambas et al., 2003; Alonso et al. 2004; 2006),
we selected galaxies in pairs adopting projected relative distance
and relative velocity thresholds: $r_{\rm p} < 100$ kpc $h^{-1}$ and
$\Delta V < 350$ km $s^{-1}$.
We found that these limits are adequate to define galaxy pairs with enhanced
star formation activity. In these works we also detected a clear
correlation between
star formation activity and the proximity to a close neighbour supporting the
physical scenario where tidal torques generated during the interactions
trigger gas inflows, feeding the star formation activity.

For the purpose of analysing the effects of close interactions, which are
known to produce the strongest effects, we selected close pair galaxies by
requiring
$r_{\rm p} < 25$ kpc $h^{-1}$ and $\Delta V < 350 $ km $s^{-1}$. With this
further restriction to closer relative separations, the effects of
interactions are largely increased as shown by Lambas et al. (2003).
Galaxy pairs are only selected within a redshift range $0.01<z<0.10$ in
order to avoid strong incompleteness at larger distances as well as
significant contributions from peculiar velocities at low redshift.
The final pair catalog in the SDSS-DR4 comprises 1607 close galaxy pairs.
Following the procedure explained by Lambas et al. (2003), we
constructed a control sample by selecting those galaxies without
a close companion within $r_{\rm p} < 100 $ kpc $h^{-1}$ and $\Delta V <
350 $ km $s^{-1}$. The control sample is also required to
have the same redshift and magnitude distribution as our close galaxy
pair catalog.

As discussed in Alonso et al. (2006), the effects of incompleteness or
aperture (e.g see also Balogh et al.2004ab) do not
introduce important bias in the galaxy pair catalogs.
Regarding incompleteness, by combining the spectroscopic and photometric
surveys, Alonso et al. (2006) estimated that the spectroscopic
catalog has only an incompleteness of $\approx$ 9.5$\%$.
Hence, neither effect is expected to have a strong impact in our close
galaxy pair catalog.

\section{The Hosts of Type 2 AGNs in Galaxy Pairs}

As a first step, we identify AGN hosts in our close pair catalog (i.e.
pairs with $r_{\rm p} < 25$ kpc $h^{-1}$ and
$\Delta V < 350 $ km $s^{-1}$) by cross-correlating it with the AGN catalog.
We found 498 galaxy pairs with one member exhibiting AGN activity.
This represent $\approx$ 30$\%$ of the total close pair catalogue.
We also detected that 108 close pairs have both members classified as AGNs
($\approx$ 7$\%$).
Then, the final SDSS-DR4 AGN-pair catalog comprises 606 pairs.

For the purpose of properly assessing the significance of the results
obtained from AGN galaxies in pairs, we defined a sample of isolated AGN
galaxies from
the SDSS-DR4 by selecting those galaxies without a companion galaxy within
$r_{\rm p} < 100$ kpc $h^{-1}$ and $\Delta V < 350 $ km $s^{-1}$, within
the same redshift range of the AGN close pair catalog.
The procedure followed to construct the sample of isolated AGN galaxies,
assures that it will have the same selection effects than AGN galaxy pair
sample, and consequently, it can be used to estimate the actual difference
between AGN galaxies in pairs and isolated AGNs,
unveiling the effect of the interactions. A similar reasoning motivated
the definition of a non-AGN galaxy sample with the same
luminosity distribution than galaxies in the AGN-pair catalog.

\begin{figure}
\includegraphics[width=90mm]{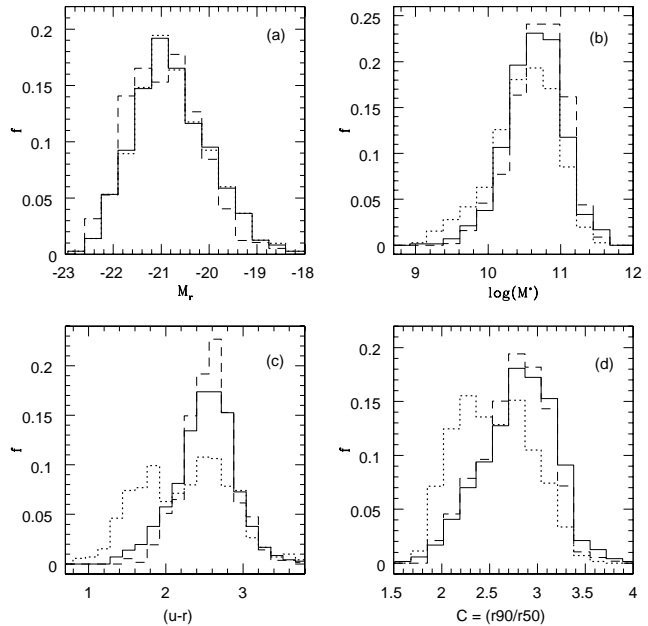}
\caption{($a$) r-band absolute magnitude, ($b$) Mass in stars , ($c$) u-r
colour and ($d$) Concentration parameter distributions . Solid lines
correspond to
AGNs in pairs, dashed lines to AGNs without companions, and dotted lines to a
control sample of non-AGN galaxies with a similar luminosity
distribution than AGNs.
}
\label{fig1}
\end{figure}

The distribution of absolute magnitude in r-band ($M_r$) and the stellar
mass ($M^*$), of AGNs in pairs, isolated AGNs,
and a control sample of non-AGN galaxies with a similar luminosity
distribution than the hosts of AGNs are shown in Fig.\ref{fig1}.
It can be seen that the luminosity distribution is similar in the three
samples, but the fraction of large stellar mass objects among AGN hosts is
larger
than that of non-AGN hosts:
the percentage of galaxies with $log(M^*) > 11$ is $~ 15.8\% $ for AGNs
in pairs, $19.7 \%$ for isolated AGNs and $9.8 \%$ for non-AGN hosts.
In fact, there are very few AGNs in galaxies with
$M^* < 10^{10}$ $M_\odot$. As it is well know, the majority of low mass
galaxies have young stellar population and a significant fraction
are currently experiencing strong star formation activity.

In Fig.\ref{fig1}$c$, we plot the ($u-r$) distributions for the same samples.
As it can be appreciated from this figure colour distribution of the control
sample of non-AGN galaxies is clearly bimodal as expected (e.g. Baldry et
al. 2003).
However AGNs, regardless of being in pairs or in isolation, have
($u-r$) colours consistent with a red host population.
Fig.\ref{fig1}$d$ shows the concentration parameter
distribution ,$C$, defined as the ratio between the radii enclosing 90 \%
and 50\% of the galaxy light in the Petrosian r-band. Recall that
Strateva et al. (2001) found that galaxies with
$C>2.6$ are mostly early-type galaxies, whereas spirals and irregulars have
$2.0<C<2.6$.
From the distributions of Fig.\ref{fig1}$d$, it can be appreciated that
AGNs reside preferentially in bulge dominated galaxies in contrast to the
control sample of non-AGN galaxies which show the expected $C$ bimodality.

\begin{figure}
\includegraphics[width=84mm]{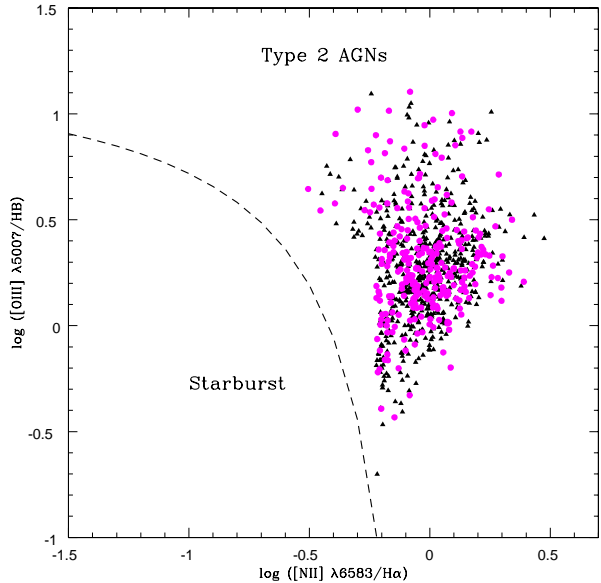}
\caption{BPT diagnostic diagram. The filled circles
correspond to AGN pair galaxies
and filled triangles to AGNs without companions.
The dashed curve represents the demarcation between starburst galaxies
and Type 2 AGNs defined by Kauffmann et al. (2003).
}
\label{OIIINII}
\end{figure}

In Fig.\ref{OIIINII} we show the BPT diagnostic diagram, plotting the ratio
$[OIII]\lambda 5007/H \beta$ versus ratio $[NII]\lambda 6583/H \alpha$ for
AGN pair galaxies and AGNs without a near companion.
As it can be appreciated, there is not an evident difference
between isolated and AGNs in pairs.

Hence, colours, morphology as measured by $C$, or luminosity, do not provide
evidence that interactions could play a role in the triggering or regulation
of the AGN activity. AGN hosts show similar distributions regardless of
the presence of a close companion. However, taking into account previous
results such as those of Strochi-Bergmann et al. (2001), a clear signal of
correlation between the AGN activity and interactions could be
detected by restricting to active star forming systems. In our previous
studies of galaxies in pairs we found that there is an important fraction of
galaxies that, although being close in projected space, do not exhibit strong
star formation activity. It is very unlikely that at these close projected
distances ($r_{\rm p} < 25$ kpc $h^{-1}$ and $\Delta V < 350 $ km
$s^{-1}$) pairs
could be strongly affected by interlopers (Alonso et al. 2004; Perez et a.
2005).
However, it is possible that, owing to particular characteristics of the
orbital parameters or internal structure, these
pairs do not experience strong tidal torques that can drive gas inflows.
Motivated by these facts, in the next Section we discuss a new
classification for close galaxy pairs based on a visual morphological
analysis
of their members.

\subsection{Classification of pair galaxies}

By using the photometric SDSS-DR4, we classified all galaxies in the close
pair catalogue taking into account the eye-ball detection of features
characteristics of interactions.
Three categories have been defined:

\begin{itemize}
\item Pairs with evidence of an ongoing merging process ($m$: merging
pairs)

\item Pairs with signs of tidal interactions but not necessarily merging
($t$:tidal pairs)

\item Pairs showing no evidence of distorted morphologies ($n$:
non-interacting pairs).

\end{itemize}

\begin{figure}
\includegraphics[width=84mm,height=120mm]{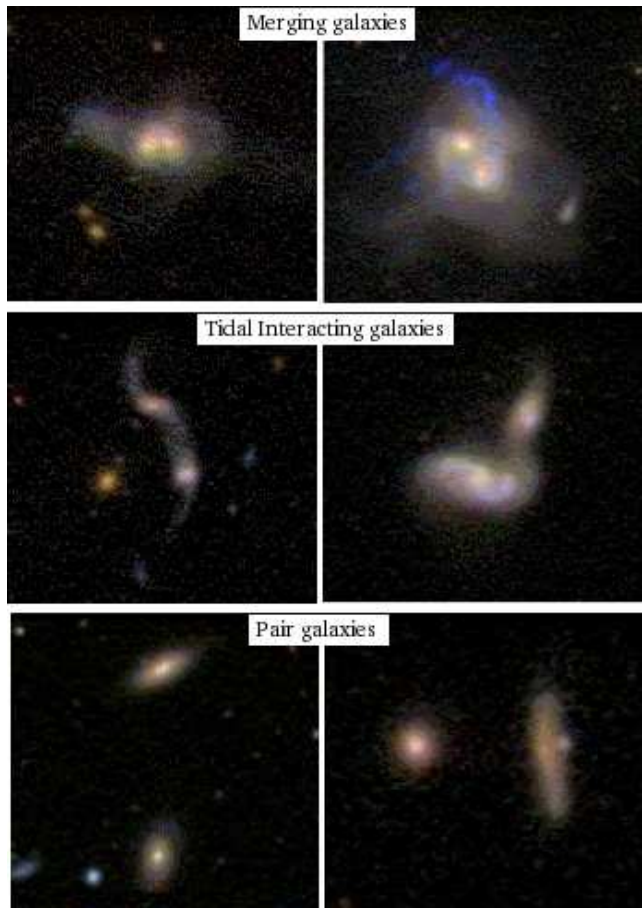}
\caption{Examples of galaxy images in close pairs with different
classification: evidence of an ongoing
merging process ,$m$, (upper panels), pair with a sign
of tidal interaction but not necessarily merging ,$t$, (medium panels)
and pair galaxies showing no evidence of distorted morphologies ,$n$,
(lower panels).
}
\label{Clasif}
\end{figure}

Fig.\ref{Clasif} shows images of typical examples of close pair galaxies
for the three different visual classification: $m$, $t$ and $n$.
Hereafter we will carry out the analysis by distinguishing them according
to these categories. Note that although the $m$ category referes to
systems in
advance stages of interactions, the two systems could be still individualized
by the SDSS survey.

\begin{table*}
\center
\caption{Percentages of AGNs}
\begin{tabular}{|c c c c c c}
Categories & Total Pairs &$m$ & $t$ & $n$ &Total Control\\
\hline
\hline
Number of close pairs &1607 & 383 & 688 & 536 & 14359  \\
\hline
$\%$ of AGNs &32\% & 31$\%$ & 32$\%$ & 28$\%$ & 23\%  \\
\hline
\end{tabular}
\end{table*}

\subsubsection{Analysis of deblending for Merging Pair Galaxies}

In order to analyse the effects of how reliable SDSS photometric
deblending procedure works in merging galaxy pairs ($m$) for which
we expect the largest uncertainties,
we have carried out suitable tests to determine independent magnitudes
for this sample.

\begin{figure}
\includegraphics[width=84mm]{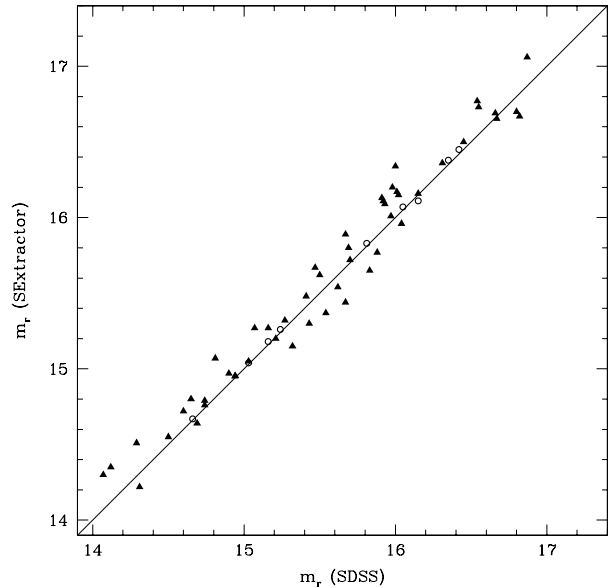}
\caption{r-band magnitud derived from SExtractor routines versus
r-band magnitud extracted from SDSS, for a sample of isolated galaxies
(open circles)
and for a subsample of close pair galaxies (triangles).
}
\label{magmag}
\end{figure}

For this purpose we have studied directly galaxy images extracted
from SDSS and we have used SExtractor routines to derive galaxy magnitudes.
We have used first, a set of isolated galaxies which serve as a control
sample for testing the photometry. For this sample,
our photometry gives r-band magnitudes with 0.02 rms difference
compared to those quoted in SDSS, indicating the consistency of our
results with SDSS magnitudes when no deblending procedure is required.
The second sample is composed of 50 merging galaxy pairs ($m$) extracted
at random from our total sample of 383 m-pairs (with typical projected separation
$r_p <$ 12 kpc $h^{-1}$).
We calculated the total magnitude of the pair
which acording to our previous analysis should be within 0.02
magnitudes compared to SDSS measurements in the case that deblending
is accurately working. We find, however, that the
deblending procedure in SDSS introduces a larger uncertainty since by
adding the quoted luminosities in SDSS, the typical difference raises
to 0.12 rms.
In Fig.\ref{magmag} we show r-band magnitud derived from SExtractor
routines versus
r-band magnitud extracted from SDSS, for a sample of isolated galaxies
and for a subsample of 50 merging pairs.
This simple and robust test gives useful estimates of the 
uncertainties introduced by the deblending procedure
adopted in SDSS which we argue are about 0.12 mag.
We also notice the lack of systematics in SDSS deblending procedures, since
the difference between our Sextractor measurements and quoted SDSS
magnitudes have an approximately zero mean.

In sum, this analysis indicates that SDSS
magnitudes can be suitably used in our analysis in the following
sections provided a rms scatter of 0.12 magnitudes is not seriously
affecting the results.

\subsubsection{Frequency of AGNs}

For the three categories defined above, we calculate the percentage of
pairs with AGNs.
As we can notice from Table 1, we found that
$\approx$ 30 $\%$ of close galaxy pairs host AGNs, regardless of the
category they belong to, and that isolated galaxies
without companions have a lower AGN frequency ($\sin 23 \%$).

\begin{figure}
\includegraphics[width=84mm]{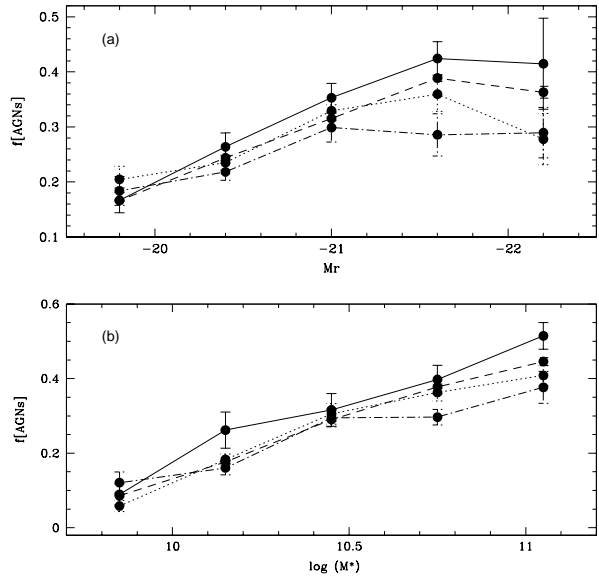}
\caption{Fraction of AGNs in merging-pair galaxies (solid lines),
in tidal pairs (dotted lines), in non-interacting pairs
(dot-dashed lines) and in galaxies without companions
(dashed lines), as a function of $M_r$ (a) and $M^*$ (b).
}
\label{fracAGN}
\end{figure}

Since AGN phenomenon is a strong function of host luminosity, we analysed the
fraction of AGNs in each of the pair categories and in the control sample,
for different stellar mass content and luminosity intervals.
In Fig.\ref{fracAGN}, we display the fraction of AGNs as a function of
r-band luminosity and stellar mass
The results indicate that the fraction of host with AGN activity
increases with luminosity in a similar fashion for AGNs in pairs or
in isolation. However, we notice that, at a given mean luminosity or mean
stellar mass, merging pairs tend to have a higher frequency of AGNs than
pairs in the two other categories or in the control sample.
Nevertheless, this excess is approximately $\approx 10\%$.

\subsubsection{Stellar population ages in AGN hosts}

The strongest discontinuity occurring at 4000 $\r{A}$ in the optical
spectrum of a galaxy arises of the accumulation of a large number of
spectral lines in a narrow wavelength region where the main contribution
to the opacity comes from ionized metals.
The break index $D_n(4000)$ (Kauffmann et al. 2002) is defined
as the ratio of the average flux density by using narrow continuum bands
(3850-3950 $\r{A}$ and 4000-4100 $\r{A}$).
The $D_n(4000)$ indicator is suitably
correlated to the mean age of the stellar population in a galaxy
and can be used to estimate the star formation rate per unit stellar mass,
$SFR/M*$, (Brinchmann et al. 2004).
The majority of star formation takes place preferentially in galaxies
with low $D_n(4000)$ values, for instance, only $12\%$ and $2\%$ of the
total SFR density are associated to galaxies with
$D_n(4000) > 1.8$ and $D_n(4000) > 2.0$ respectively.

\begin{figure}
\includegraphics[width=84mm]{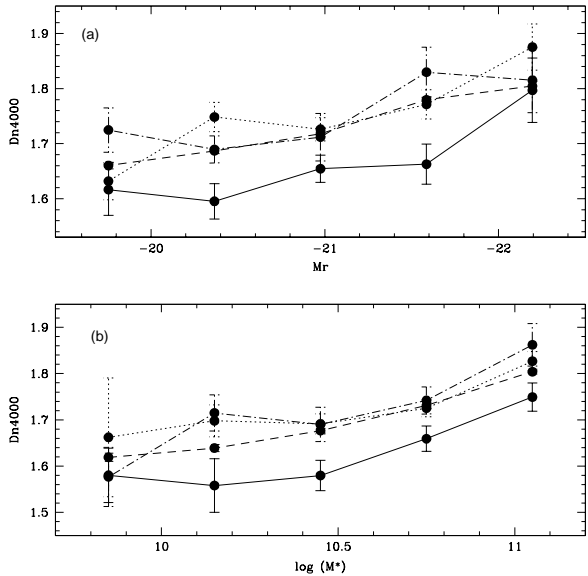}
\caption{$D_n(4000)$ as a function of (a) $M_r$ and (b) $M^*$.
The different types of lines indicate AGNs in merging-pair galaxies (solid),
AGNs in tidal-pairs (dotted), AGNs in non-interacting pairs
(dot-dashed) and AGNs in galaxies without companions
(dashed).
}
\label{DnMrMst}
\end{figure}

We have analysed the behaviour of this parameter as a function of
stellar mass and luminosity for AGNs in pairs and in isolation.
The results are given in Fig.\ref{DnMrMst} where it can be seen that
AGNs in merging pairs have significantly lower values of $D_n(4000)$. This
finding suggests that these galaxies have experienced more recent
episodes of star formation as expected since they are effectively in an
ongoing merger event.

The trends shown in Fig.\ref{fracAGN} and Fig.\ref{DnMrMst} support the
scenario where interactions could drive both star formation and AGN activity
via the triggering of gas inflows to the central region.
The fact that we are actually detecting the strongest signal for merging
pairs could be indicating that fusions could be more efficient than tidal
torques produced during the orbital decay to feed both
AGN and star formation activity.

\section{Properties of AGNs in interacting Galaxies: O[III] Luminosity and
Black Hole Mass}

As a tracer of the AGN activity, we focus here on the luminosity of the
[OIII]$\lambda$5007 line, calculated by Kauffmann et al. (2003) in SDSS-DR4
AGN galaxies.
Although this line can be excited by massive stars as
well as by AGNs, it is known to be relatively weak in metal-rich,
star-forming galaxies.
The [OIII] line also has the advantage of being strong and easily detected
in most galaxies.
It should be noticed that the narrow-line emission
is likely to be affected by dust within the host galaxy (Kauffmann 2003).
Thus it is important to correct the L[OIII] luminosities for the effects
of extinction.
For AGNs in SDSS, Kauffmann et al. (2003) measured the extinction
using the Balmer decrement finding that the best approximation for a dust
correction to L[OIII] is based on the ratio H$\alpha$/H$\beta$.
We will considered the luminosity $L[OIII]= 10^{6.5} L_{\odot}$ as a limit
between weak and powerful AGNs. A weak L[OIII] emission indicates that the
black hole
is not rapidly growing (Heckman et al. 2004).

We estimated black holes masses, $M_{BH}$, for both AGNs with and without
companion, by using the observed correlation between $M_{BH}$ and the
bulge velocity dispersion $\sigma_*$ (Tremaine et al. 2002)

\begin{equation}
logM_{BH} = 8.13 + 4.02 log(\sigma_*/ 200)
\end{equation}

We restrict this analysis to AGN galaxies with $\sigma_*$ $>$ 70 km
$s^{-1}$,
corresponding to log$M_{BH}$ $>$ 6.3, because the instrumental resolution
of SDSS spectra is $\sigma_*$ $\approx$ 60 to 70 km $s^{-1}$.
The aperture corrections to the stellar velocity dispersions are small in
early-type galaxies in SDSS (Bernardi et al. 2003), so we do not apply
such corrections.

\begin{figure}
\includegraphics[width=84mm]{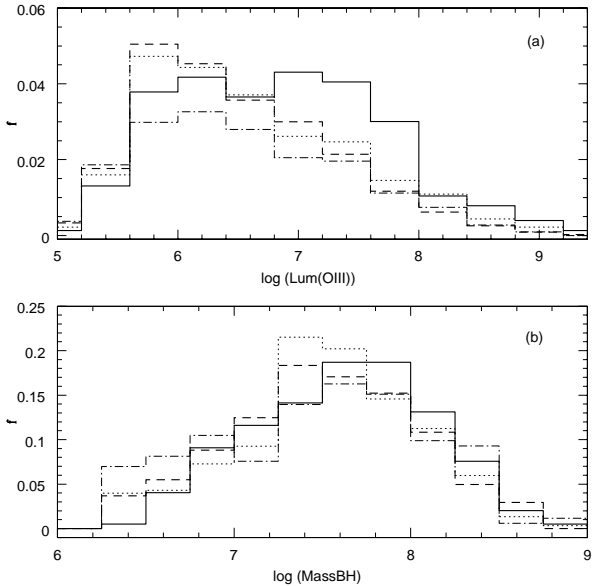}
\caption{Distributions of L[OIII] ($a$) and Black Hole Mass ($b$) in
AGNs merging-pair galaxies (solid lines),
AGNs tidal-pairs (dotted lines), AGNs non-interacting pairs
(dot-dashed lines) and AGNs galaxies without companions
(dashed lines).
}
\label{histLOIII}
\end{figure}

In Fig.\ref{histLOIII}a, we show the distribution of $L[OIII]$ for AGNs in
galaxies in merging pairs, tidal-pairs, non-interacting pairs
and without a close companions.
As it can be clearly appreciated, AGNs in pairs with evidence of
ongoing merging processes ($m$) show larger luminosities $L[OIII]$ than
AGNs in isolation or in the other two pair categories. In fact,
while $62.4\%$ of AGNs in merging pairs have $L[OIII]> 10^{6.5} L_{\odot}$,
only $43.3\%$ of isolated AGNs show such a powerful activity.
Note that the other two pair categories have AGN luminosity distributions
similar to that of isolated AGNs.

In Fig.\ref{histLOIII}b , we show the distribution of
black holes masses, $M_{BH}$, for AGNs grouped similarly to
Fig.\ref{histLOIII}a.
It can be appreciated in this figure that although there is a trend for
$M_{BH}$ to be
systematically larger for AGNs in merging pairs, the difference is not as
significant as the signal detected for $L[OIII]$, being only of less than
0.15 dex.

\begin{figure}
\includegraphics[width=90mm]{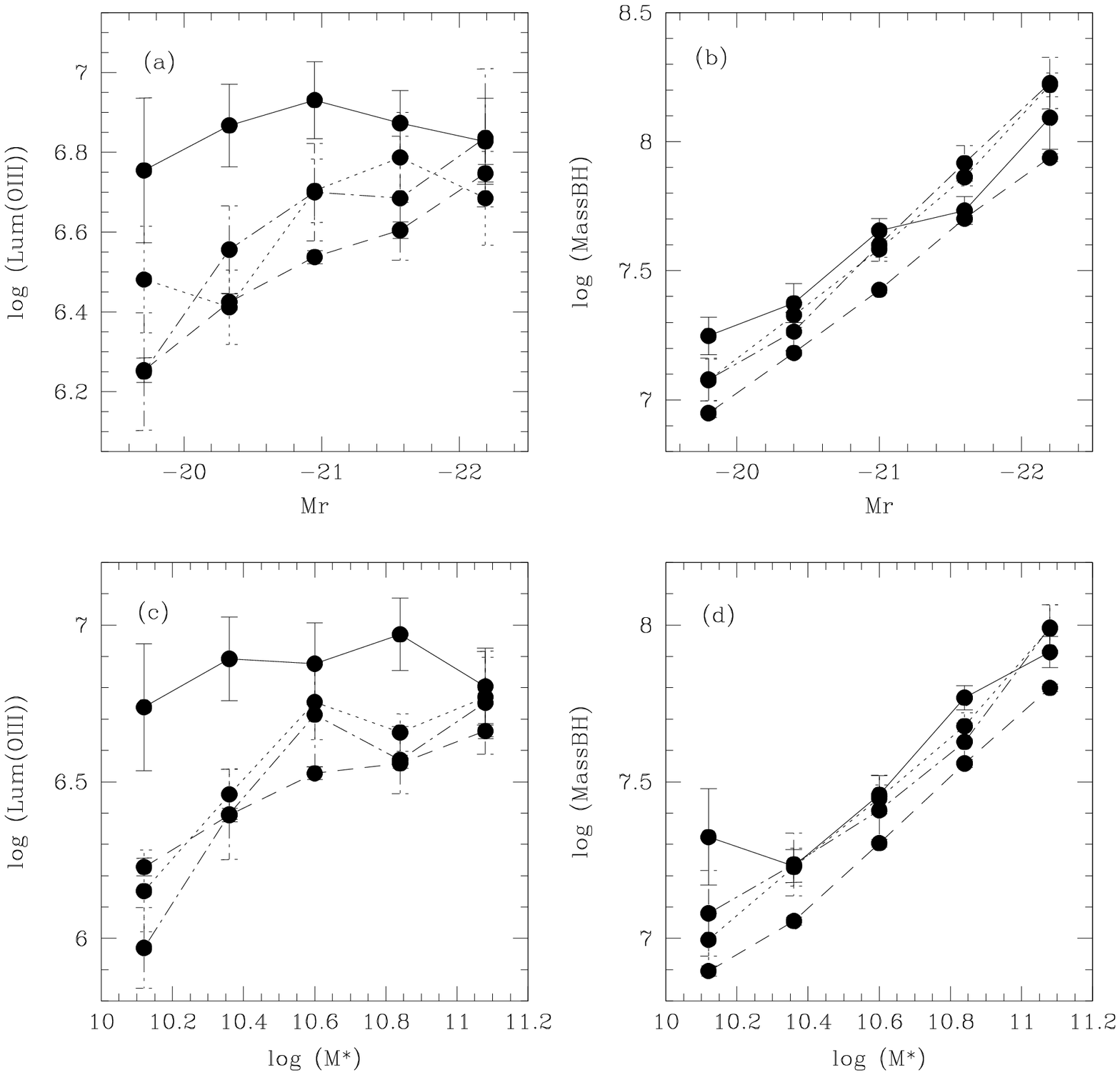}
\caption{L[OIII] (a,b) and $M_{BH}$ (c,d) as a function of $M_r$ and
$M^*$ for
AGNs in merging pairs (solid lines),
tidal pairs (dotted lines), non-interacting pairs
(dot-dashed lines) and in AGNs galaxies without close companions
(dashed lines).
}
\label{LumOIIIM}
\end{figure}

It is interesting to further investigate the strength of AGNs as a
function of host luminosity and stellar mass.
Therefore, we calculated the mean $L[OIII]$ as a function of r-band
luminosity and stellar mass content of the corresponding host galaxies.
These relations have been estimated for AGNs in the three pair categories
($m$, $t$, $n$) and for isolated AGNs.
From the results shown in Fig.\ref{LumOIIIM}a, it can be clearly
appreciated that, in general,
most luminous hosts show the highest $L[OIII]$.
We can also see that AGNs in merging pairs show higher $L[OIII]$,
regardless of the luminosity and stellar mass of the host galaxy,
indicating that the black hole activity is stronger for AGNs in advance
stages of interactions.
In fact, for host galaxies with $M^*> 10^{10.5} M_{\odot} $, the mean
$L[OIII]$ of merging pairs is larger by half an order of magnitude than
the mean
$L[OIII]$ of the rest of the samples.
A similar estimation for $M_{BH}$
confirms the trend found in Fig.$\ref{histLOIII}$b. Black hole masses are
slightly larger for AGNs in merging pairs than for the rest of the
categories.

\begin{figure}
\includegraphics[width=75mm,height=140mm]{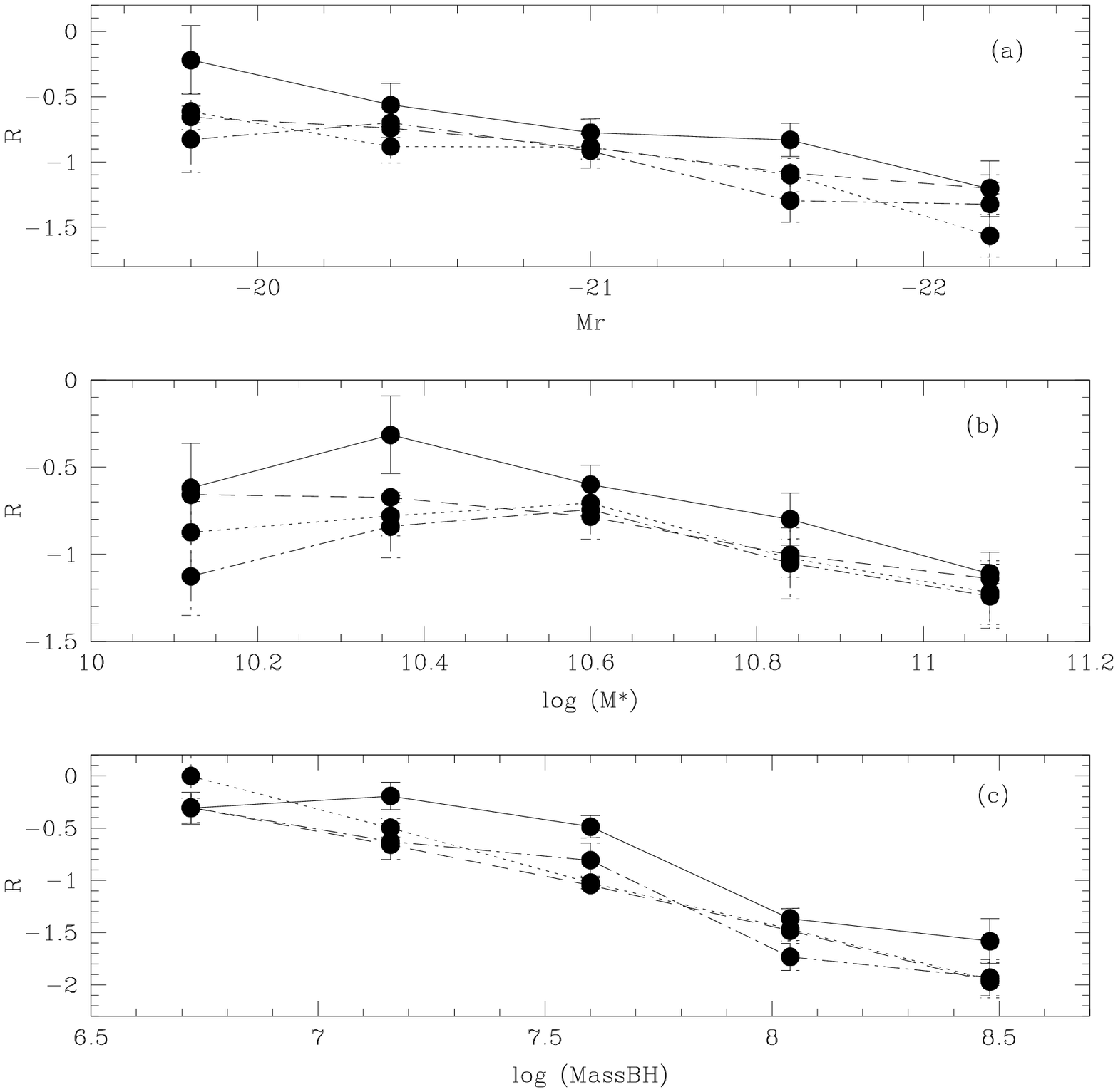}
\caption{Mean $\cal R$ = log($L[OIII]$/$M_{BH}$) as a function of $M^*$
($a$), $M_r$ ($b$)
and $M_{BH}$ ($c$) for AGNs in merging pairs (solid lines),
tidal pairs (dotted lines), non-interacting pairs
(dot-dashed lines) and AGNs galaxies without close companions
(dashed lines).
}
\label{LumOIIIBHM}
\end{figure}

The ratio of [OIII] luminosity to black hole mass ($\cal R$ =
log($L[OIII]$/$M_{BH}$))
provides a useful measure of the accretion rate onto a black hole (Heckman
et al. 2004).
In our calculations we have not converted the $L[OIII]$ to volume-averaged
luminosities but by using the relations displayed in Fig. 2 of
Heckman et al. (2004) it is possible to infer an accretion time for AGNs
in pairs.
In Fig.\ref{LumOIIIBHM} we show $\cal R$ as a function of
host luminosity (a), stellar mass (b) and BH mass (c),
for AGNs in the three defined pair categories ($m$, $t$, $n$) and in
isolated hosts.
In agreement with the previous results, we found that black holes in
smaller or fainter systems are
more active than black holes in larger or brighter systems.
Heckman et al. (2004) concludes that the most rapidly
growing black holes are those with $M_{BH} < {\rm few \times 10^{7}
M_{\odot}}$ with an implied growth time $\sim$ twice the age of the universe.
From Fig.$\ref{LumOIIIBHM}$c we can appreciate that $M_{BH} < {\rm few \times
10^{7} M_{\odot}}$ have significantly higher accretion rates than larger
ones.

The relations shown in Fig.\ref{LumOIIIBHM} indicate that, although in
general the smallest black holes are the ones that exhibit the highest
rates of accretion, AGNs in merging pairs host the most rapidly growing
black holes.
In fact, AGNs in merging pairs have the most powerful black holes and the
ones that are growing faster, besides hosting
stellar populations which show clear signal of being importantely
rejuvenated by recent starbursts.

\begin{figure}
\includegraphics[width=90mm]{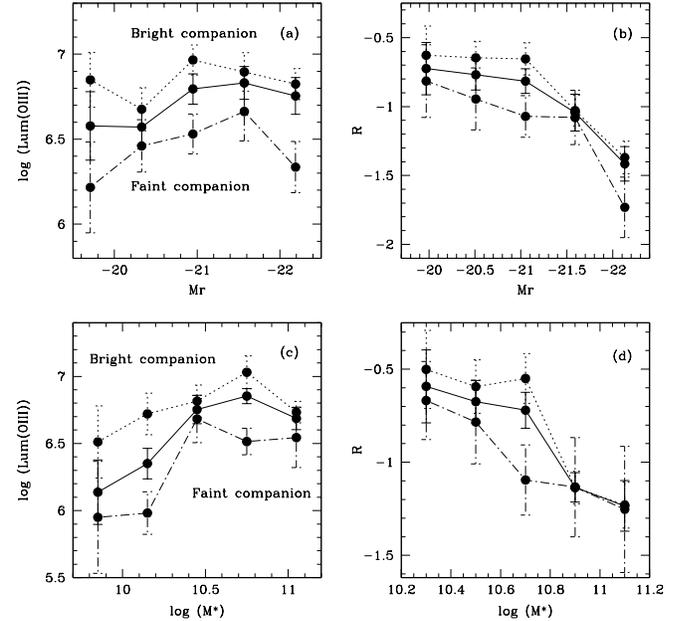}
\caption{Mean $Lum[OIII]$ ($a$, $c$) and $\cal R =L[OIII]$/$M_{BH}$ ($b$,
$d$)
as a function of $M_r$ and $M^*$ for AGNs in pairs (solid lines), AGN
galaxy with a bright pair companion
($Mr<-20.0$) (dotted line), and AGN with a faint pair companion
($Mr>-20.0$) (dashed line).
}
\label{OIIIMcomp}
\end{figure}

Finally, we have also investigated the role played by the AGN companion in
powering the AGN activity.
This was accomplished by calculating, for each AGN in pairs, the OIII
luminosity and $L[OIII]$/$M_{BH}$
as a function of its r-band luminosity and stellar mass content,
considering separately pairs with a bright
and a faint companion. For that purpose, we adopted $M_r =-20$ as the
magnitude threshold.
The results shown in Fig.\ref{OIIIMcomp} clearly indicate
that AGN activity in hosts with bright companions is significantly
enhanced with respect to that of AGNs with faint companions, being
both the AGN power and accretion rate significantly larger.
This finding provides evidences that nuclear activity
is not only affected by the presence of a very close companion
but that the relative mass (or luminosity) is
also an important issue to take into account.

\section{Summary and Conclusions}

We have performed a statistical analysis of both, host characteristics
and nuclear activity, of AGNs in pairs and without close companions. Our
study is based on the sample of AGNs derived from the SDSS-DR4 release by
Kauffmann et al. (2003) and the pair galaxies obtained from the same data
release.
We have complemented the SDSS-DR4 data with
the addition of a eye-ball classification of images of 1607 close pairs
($r_p<$25 kpc $h^{-1}$, $\Delta V<$ 350 km $s^{-1}$) according to the
evidence of interaction through distorted morphologies and tidal features.
Also, we have performed a photometric analysis to assess the reliability
of the SDSS
magnitude deblending procedure in merging pairs, finding negligible
systematic
effects.

Our main results are:

1- The fraction of AGNs as a function of r-band luminosity or stellar
mass is larger for merging pairs.
However, this is a relative small effect accounting for an
increase of less $\sim 10\%$
due to tidal interactions.

2- AGN hosts are redder and with a larger concentration morphological
index than non-AGN galaxies. This effect is similar for AGN hosts in pairs
or in
isolation.

3- We find that the locus of AGNs with and without close companions in the
BPT diagram are consistent with each other.

4- At a given host r-band luminosity or stellar mass, we find the AGN OIII
luminosity to be enhanced for AGNs with strong interaction features
(i.e. merging pairs). We found that while $62\%$ of AGNs in merging pairs
have $Lum[OIII] > 10^{6.5}$ this percentage decreases
to $43\%$ in the case of isolated AGNs.

5- Estimations of the mean accretion rates onto the black holes of AGNs in
pairs in the three defined categories and in isolation indicate that
AGNs in merging pairs have more active black holes than other AGNs,
at a given r-band luminosity or stellar mass content.
We estimated than $27 \%$ of AGNs in merging pairs have $Lum[OIII] >
10^{6.5}$ and $M_{BH} < 3 \times 10^7 M_{\odot}$ and can be considered as
being feeding
their black holes with high efficiency (Heckmann et al. 2003). In the case
of isolated
AGNs (selected to have the same , this
percentage decreases to $21\%$. On the other hand, all AGNs in merging pairs
have $Dn4000 < 1.8$ which indicates that they also have active star formation
formation. Interestingly we also found that $17\%$ of isolated AGNs have
powerful emission, $M_{BH} < 3 \times 10^7 M_{\odot}$ and have also
experienced important star formation activity in recent past with $Dn4000
< 1.8$. These trends open the possibility that isolated AGNs with active
black holes have experienced a recent merger. It is also possible that
secular
evolution could be responsible of feeding the black hole and the star
formation activity if a gaseous disk is present. More detail data are
needed to draw a conclusion on this point.

6- We find that the luminosity (or mass in stars) of the companion galaxy
is a key parameter to determine the activity of the black hole.
Both the OIII luminosity and the accretion rates $L[OIII]/M_{BH}$ for AGNs
with bright close companions are significantly larger than for those of
AGNs with faint
neighbours.

\section*{Acknowledgements}

We thank Guinevere Kauffmann for useful comments.

This work was partially supported by the European Union's ALFA-II program,
through LENAC, the Latin American European Network for Astrophysics and
Cosmology and Conicet from Argentina.

Funding for the creation and distribution of the SDSS Archive has been
provided
 by the Alfred P. Sloan Foundation, the Participating Institutions, the
National
Aeronautics and Space Administration, the National Science Foundation, the
U.S.
Department of Energy, the Japanese Monbukagakusho, and the Max Planck
Society.
The SDSS Web site is http://www.sdss.org/.

The SDSS is managed by the Astrophysical Research Consortium (ARC) for the
Participating Institutions. The Participating Institutions are The
University of
Chicago, Fermilab, the Institute for Advanced Study, the Japan
Participation Group,
The Johns Hopkins University, the Korean Scientist Group, Los Alamos National
Laboratory, the Max-Planck-Institute for Astronomy (MPIA), the
Max-Planck-Institute
for Astrophysics (MPA), New Mexico State University, University of
Pittsburgh,
University of Portsmouth, Princeton University, the United States Naval
Observatory, and the University of Washington.


\begin{thebibliography}{}

\bibitem{}
Alonso, M.S., Tissera, P.B., Coldwell, G. \& Lambas, D.G., 2004, MNRAS,
352, 1088.

\bibitem{}
Alonso, M.S., Lambas, D.G., Tissera, P.B. \& Coldwell, G., 2006, MNRAS,
367, 1029.

\bibitem{}
Antonucci, R., 1993, ARA\&A, 31, 473.

\bibitem{}
Baldry, I. K.; Glazebrook, K.; Brinkmann, J.; Ivezic, Z.; Lupton, R.
H.;
Nichol, R. C.;
Szalay, A. S., 2004, ApJ, 600, 681.

\bibitem{}
Baldwin, J. A., Phillips, M. M. \& Terlevich, R., 1981, PASP, 93, 5.

\bibitem{}
Balogh M., Baldry I. K., Nichol R., Miller C., Bower R., Glazebrook K.,
2004,
ApJ Letters, 615, 101.

\bibitem{}
Balogh M., et al., 2004, MNRAS 348, 1355.

\bibitem{}
Barnes, J. E. \& Hernquist, L., 1992, ARA\&A, 30, 705.

\bibitem{}
Barnes, J. E. \& Hernquist, L., 1996, ApJ, 471, 115.

\bibitem{}
Barrow, J. D., Bhavsar, S. P., \& Sonoda, B. H., 1984, MNRAS, 210, 19p.

\bibitem{}
Begelman, M. C., Blandford, R. D., \& Rees, M. J., 1980, Nature, 287,
307.

\bibitem{}
Bernardi, M. \etal, 2003, AJ, 125, 1817.

\bibitem{}
Brinchmann, J., Charlot, S., White, S. D. M., Tremonti, C., Kauffmann,
G.,
Heckman, T.
\& Brinckmann, J., MNRAS, 2004, 351, 1151.

\bibitem{}
Coldwell, G. V. \& Lambas, D. G., 2006, MNRAS, in press.

\bibitem{}
Gunn, J. E., 1979, in Active Galactic Nuclei, ed. C. Hazard \& S.
Milton
(Cambridge: Cambridge Univ. Press), 213.

\bibitem{}
Dahari, O., 1985, ApJS, 57, 643.

\bibitem{}
Fuentes-Williams, T. \& Stocke, J.T., 1988, AJ, 96, 1235.

\bibitem{}
Heckman, T. M.; Kauffmann, G.; Brinchmann, J.; Charlot, S.; Tremonti,
C.;
White, S. D. M.,
2004, ApJ, 613, 109.

\bibitem{}
Kauffmann, G., \etal, 2003, MNRAS, 341, 33.

\bibitem{}
Kauffmann, G., \etal, 2003, MNRAS, 346, 1055.

\bibitem{}
Keel, W. C. 1996, AJ, 111, 696.

\bibitem{}
Kelm, B., Focardi, P. \& Zitelli, V., 2004, A\&A, 418. 25.

\bibitem{}
Lambas, D.G., Tissera, P.B., Alonso, M.S. \& Coldwell, G., 2003, MNRAS,
346, 1189.

\bibitem{}
Martinet, L., 1995, FCPh, 15, 341.

\bibitem{}
Mihos J. C. \& Hernquist L., 1996, ApJ, 464, 641.

\bibitem[Perez]{}
P\'erez M. J., Tissera P., Lambas D. G., Scannapieco C., 2006, A\&A,
449, 23.

\bibitem{}
Sanders, D.B., Soifer, B.T., Elias, J.H., Madore, B.F., Matthews, K.,
Neugebauer, G.
\& Scoville, N., 1988, ApJ, 325, 74.

\bibitem{}
Schmitt, H. R., 2001, AJ, 122, 2243.

\bibitem{}
Schwartz, M. P., 1981, ApJ, 247,77.

\bibitem{}
Shlosman, I., Begelman, M. C. \& Frank, J., 1990, Nature, 345, 679.

\bibitem{}
Storchi-Bergmann, T.; González Delgado, R. M.; Schmitt, H. R.; Cid
Fernandes, R.;
Heckman, T., 2001, ApJ, 559, 147.

\bibitem{}
Strateva, I. \etal, 2001, AJ, 122, 1861.

\bibitem{}
Strauss, M. \etal, 2002, AJ, 124, 1810.

\bibitem{}
Toomre, A. \& Toomre, J., 1972, AJ, 178, 623.

\bibitem{}
Tremaine, S. \etal, 2002, ApJ, 574, 740.

\bibitem{}
York D. O. et al., 2000, ApJ, 120, 1579.

\end{thebibliography}
\end{document}